# From Double to Triple Burden: Gender Stratification in the Latin American Data Annotation Gig Economy

Lauren Benjamin Mushro, PhD

*Universitat de Vic – ELISAVA / Sapien AI*

# From Double to Triple Burden: Gender Stratification in the Latin American Data Annotation Gig Economy

**Word Count:** 6242

## Abstract

This paper examines gender stratification in the Latin American data annotation gig economy, with a particular focus on the "triple burden" shouldered by women: unpaid care responsibilities, economic precarity, and the volatility of platform-mediated labor. Data annotation, once lauded as a democratizing force within the global gig economy, has evolved into a segmented labor market characterized by low wages, limited protections, and unequal access to higher-skilled annotation tasks. Drawing on an exploratory survey of 30 Latin American data annotators, supplemented by qualitative accounts and comparative secondary literature, this study situates female annotators within broader debates in labor economics, including segmentation theory, monopsony power in platform labor, and the reserve army of labor. Findings indicate that women are disproportionately drawn into annotation due to caregiving obligations and political-economic instability in countries such as Venezuela, Colombia, and Peru. Respondents highlight low pay, irregular access to tasks, and lack of benefits as central challenges, while also expressing ambivalence about whether their work is valued relative to male counterparts. By framing annotation as both a gendered survival strategy and a critical input in the global artificial intelligence supply chain, this paper argues for the recognition of annotation as skilled labor and for regulatory interventions that address platform accountability, wage suppression, and regional inequalities.

## Introduction

María, a 34-year-old mother of two from Caracas, Venezuela, has worked on and off for Remotasks, a subsidiary of Scale AI, for half a year. She typically begins her day at 4 a.m., logging into Remotasks and hoping to secure a data labeling project. Sometimes, if she logs on even 15 minutes too late, all tasks have already been assigned or taken. If she does claim a task, payment is uncertain, and hours later she may still have earned nothing, depending on the reviews she receives on her annotations. Her days continue like this: between caring for her child and elderly mother, she checks her task dashboard repeatedly, only to refresh to no new tasks. María finds herself in what we will call the flexibility paradox: annotation firms often market click work as incredibly flexible and autonomous, yet at the same time, workers are tethered to their dashboards, waiting for tasks that never come due to labor surplus, sinking further into precarity rather than emerging from it. María's experience reflects a growing pattern among women in Latin America's gig economy, which is the triple burden of unpaid care, precarious income, and the volatility of digital labor markets.

Concerns surrounding the gig economy, defined as the labor market of short-term contracts or freelance workers facilitated by digital platforms, and the broader "future of work" debate are not entirely new. Scholars have documented the shifting labor market since the 2010s (Berg, 2019; De Stefano, 2015; Dubal, 2019; Frey & Osborne, 2013; Graham, 2018), with early debates emphasizing platforms like Uber, Lyft, Upwork, and Fiverr. The International Network on Digital Labor meets yearly to discuss the impacts of platform mediated work on workers' rights. Only more recently has focus shifted to a specific subsector of the gig work economy: data annotation, or the repetitive click work and labeling of data points to inform Large Language Models (LLMs), facilitated by digital platforms like MTurk, Scale, and Appen. Data annotation can take many forms: semantic segmentation, image annotation, bounding boxes, automated speech recognition, timestamping, cuboids, and sentiment analysis, among many others. For certain annotation firms, data annotation has become a form of disguised wage labor, where task payouts are determined by typical earnings per hour and therefore suppressed.

While originally celebrated as a democratizer of labor supply, digital click work has evolved into a form of platform monopsony, where a few dominant firms exert disproportionate control over workers' access to wages, hours, psychological support (for example, from tasks involving exposure to violent and extreme images), and dispute resolution. Far from creating an open opportunity market, annotation has reinforced dynamics of labor surplus extraction and wage suppression.

Much of the early 2020s emphasized the sheer amount of data and data annotators as the most important aspect of creating a functional LLM. In 2025, while we still see this demand, the rise of products like DeepSeek have demonstrated the need for less quantity and more quality, with the ability to train models on small, high-quality datasets through reinforcement learning alone (DeepSeek, 2025). In the race with China to reach AGI, methods for training models have transformed drastically. While in the past, machine learning researchers relied on large datasets for training models, more recently, fine-tuning through inference on Nvidia's H20 chip has become a much more popular method that trades higher quantities of data for higher quality data (Center for Strategic and International Studies, 2025).

This has left data annotators more vulnerable than ever before in an already precarious field. Without the benefits of full-time employees, like paid time off, psychological resources, or even access to basic resources like the bathroom in large-scale data farms, these workers now find a dearth of low-skilled opportunities as companies consolidate under loose antitrust regulation (see Scale AI's partial acquisition) and are forced to compete with a larger workforce than ever before.

Even further than this precarity, studies show an uneven distribution of gig work across the globe, with the majority originating from what is often referred to as "The Global South," or more appropriately, the "Majority World," referring to "countries located in Africa, Asia, Latin

America, and the Caribbean," with centralization in the following countries: Venezuela, Colombia, Nigeria, Vietnam, India, and the Philippines (Foreign Analysis, 2025; Sapien geodistribution, 2025).

This paper aims to address the overwhelming number of data annotators originating from Latin America, with a focus on female data annotators, who often face the most difficult conditions for part-time work, simultaneously juggling 1) caregiving, 2) economic precarity, and 3) platform labor dynamics. While several female annotators in Latin America who took part in this study's survey do not report gender-based discrimination in their work, this absence of perception does not negate the reality of gendered labor dynamics that are likely to unfold in the coming years. For these women, gig work is not only an individual choice but part of household-level survival strategies that distribute burden unevenly on women by shifting unpaid, care, and freelance labor onto women. The structure itself of data annotation work disproportionately draws in women precisely because of gendered economic pressures and caregiving responsibilities.

**Survey and Methodology**

This paper draws on a proprietary survey conducted between May and August 2025 with 30 Latin American data annotators through online platform communities. While limited in scale, the findings align with broader trends documented by the International Labour Organization and Fairwork Institute, and they provide critical micro-level evidence of how gender stratification manifests in platform-mediated labor markets in Latin America, an area with little to no statistical resources on the state of data annotation.

This survey was distributed through formal communities like MTurk, Qualtrics, and other informal annotator communities like Whatsapp and Signal, using snowball sampling across Chile, Colombia, México, and Venezuela. Key limitations include overrepresentation of workers with reliable internet and electricity access, and self-selection bias toward respondents with predisposed willingness to discuss working conditions. Similarly, almost all respondents on MTurk and Qualtrics self-identified as male, which aligns with reports that indicate higher activities from men on these types of annotation platforms (Litman, 2020). Despite these limits, the data provide valuable insights into the demographics, motivations, and challenges of female data annotators in Latin America.

**Demographics and Conditions of Female Data Annotators in Latin America**

According to the International Labor Organization's (ILO) 2021 Workforce report, as a result of the pandemic, global female employment declined by 5%, compared to the 3.9% decline for men, and nearly "90% of women who have lost their jobs have subsequently left the workforce, a far higher rate of inactivity than for men" (ILO, 2021). The ILO's study surveyed 181 freelance platforms, 46 of which were dedicated to microtasks like data annotation.

Though the data annotation field is lauded as a "planetary labor market" due to the ability to perform work from any corner of the world, access to work and treatment on the job is often directly tied to demographics like geographical location, gender, education level, time zone, and internet access. Because of this, "a key feature of cloudwork is the uneven geographies and regional inequalities that permeate these new work arrangements" (Fairwork Foundation, 2025, p. 6).

While there is a more established literature around female data annotators in India, Kenya, and Nigeria (Ghosh, 2023; Gray & Suri, 2019), research around conditions for female workers in Latin America remains sparse. Preexisting studies on India uncover that a large portion of women working as annotators are "frequently forced to surrender their income to their husbands" (Williams et al., 2025). Reports of data farms, or factories where hundreds of annotators serve long shifts conducting click work, similar to sweatshops, in Kenya and Nigeria demonstrate that companies like OpenAI paid workers less than $2 an hour to enhance their chatbot's safety mechanisms. Not only is this poor treatment of gig workers a form of modern enslavement and wage suppression, but it is also a threat to the very integrity of the chatbots that society uses day in and day out. As we have seen for decades, garbage data inputs create garbage data outputs: meaning poor factory conditions, regardless of the product, create poor product outcomes (Bloomberg, 2023).

Treatment of workers is not the only consideration in the effect on annotation output. Similarly, specific demographic groups tend to be hired more frequently for certain higher-skill tasks. For example, in Kenya and Nigeria, certain types of data annotation, like 3D annotation, cloud point annotation, 2D to 3D mapping, and other skilled tasks, tend to be carried out by Business Process Outsourcing companies that are male-dominated (Sapien demographics) because of ease of access to education and financial support to learn new skills.

One other important factor of the availability of gig work is dependent on geography and linguistic hierarchies. While taggers across the globe often report lack of availability of tasks, even at a basic skill level, linguistic hierarchies also play into availability of task work on platforms. Customer demand demonstrates a strong preference for European Spanish, according to the Real Academia Española, over any Latin American dialects, which can vary greatly from Iberian linguistics. Not only this, but Spanish audio recordings from the Iberian Peninsula can cover over three times the payout rate of Spanish audio recordings from countries across Latin America. *Freelancer*, a preeminent Spanish-speaking platform for online freelance projects across web, copyediting, and design, found that workers from Latin America were "42% less likely to secure bids posted by employers in Spain" with a 16% lower wage increment than workers directly located in Spain, demonstrating not only poor treatment but open discrimination incorporated into the very practice of these platforms (Williams et al., 2025).

Latin America presents a distinct case in comparison with India, Kenya, and Nigeria given that much of data annotation labor has consolidated in countries experiencing continuing economic and political instability: Chile's bus fare strikes, snap elections in Peru, and Maduro's economic crisis. Further, while formal dollarization only exists in countries like El Salvador, informal dollarization proliferates in countries like Venezuela to cope with hyperinflation and control policies after the Central Bank of Venezuela stopped publishing official figures on inflation (Globalization Guide, 2025; Reuters, 2018). With the minimum wage in Venezuela at $2.40 USD per hour, gig work provides an enticing opportunity to supplement a low local wage.

Reports originating from Latin America, including this study's proprietary report, indicate the highest concentration of gig workers come from Venezuela, Colombia, Chile, and Peru, largely due to the lack of reliable, available work in home countries due to political and economic turmoil, which not only pushes workers to seek out gig work but also to emigrate from Venezuela to neighboring countries like Colombia and Peru. These dynamics in Venezuela continue to push nationals to neighboring countries like Peru, Colombia, Ecuador, and Chile, which accounts for the high survey participation rates from those countries (International Organization for Migration, 2024). In comparison with annotators in India, who have access to limited but distinct welfare resources like maternity benefits, social pensions, and housing subsidies, as well as better access to internet, the lack of infrastructure in Venezuela and neighboring countries due to widespread corruption and political upheaval creates an environment in which basic welfare resources are not readily accessible (Rahman, 2025). Because of this, a large number of our survey respondents indicate that they work two or more jobs, including their annotation gig work.

A survey of *Toloka* gig workers in Latin America demonstrated that 59% of platform workers were single, looking to supplement their main income. Patterns emerged specifically for women data annotators. Our study indicated that 68.8% of survey respondents identified as female, indicating a slightly larger portion of women conducting gig work and/or a larger portion of women open to discussing their work as annotators. In annotator circles, for communication, "women tended to experiment with a wider range of technologies for this purpose, while men primarily relied on WhatsApp" (Williams et al., 2025, p. 9). Similarly, annotators cited 'financial independence' as the main driver for completing gig work, allowing them to achieve independent, but limited, upward mobility. While workers identified that "this type of work [crowd work] does not discriminate against genders," the work status of our survey respondents indicates a need to supplement main income sources with invisible sources like caretaking, cleaning, and gig working, types of work that largely fall to women and migrants.

Respondents cited specific platforms they tend to work on including *Remotasks* and *Outlier* (Scale AI's third-party vendors), *Appen, Telus International, TransPerfect,* and *Populii*. A substantial 71.4% of our survey respondents indicated they work two or more jobs, with 50% of respondents flagging that they only sometimes or infrequently receive payment for their work.

Female respondents cite main challenges, in order of frequency of mention, as: low pay and lack of benefits, infrequent availability of work, difficulty with time management, management of anxiety, and long hours of sedentary work. These concerns do not stand in isolation but form an interlinked system of burdens: economic precarity, temporal fragmentation, and embodied or psychological strain. Our respondents illustrate the "triple burden" in practice – gendered disadvantages specific to women in the data annotation field that compound existing inequalities.

Even further than simple disadvantaged patterns of work in data annotation as a woman, political factors and instability that push annotators from Venezuela, Colombia, Chile, and Peru to turn to gig work are often first and foremost absorbed by women, who are continuously failed by welfare systems in basic ways. In 2020, in the early years of the humanitarian crisis in Venezuela, the Centro de Justicia y Paz released a report that detailed the amplification of women's vulnerability through weakened institutions and governance, impoverishment, lack of access to food and medicine, and rising forced migration, among other factors. In her book *Invisible Women: Data Bias in a World Designed for Men*, Criado Pérez acknowledges the difficult fact that crises, economic, political, and environmental, magnify existing gender inequalities, with women being the first to lose access to food, healthcare, and social protection: "Disaster response is designed for men, and women pay the price" (Criado Pérez, 2019). Pressingly, the lack of access to food and medicine necessitates that female caretakers take on multiple jobs, one of the most popular of which in recent months is data annotation.

Though AI was invented over five decades ago by John McCarthy and Alan Turing, the widespread adoption of Artificial Intelligence and understanding of data annotation itself is relatively recent, and gig work only became part of the collective consciousness of workers in Latin America in the 2020s, around heightening political and economic upheaval. When compared with India, which has an established literature around gig work and the treatment of female data annotators, a class consciousness around invisible laborers, specifically female data annotators, has not yet formed in Latin America.

One annotator I spoke with stated that data annotation is still "muy poco conocido en Venezuela. En mi entorno poca gente tiene idea del tipo de trabajo de anotación que hago. Siempre se sorprenden cuando les cuento" [very little known in Venezuela. In my environment, few people have an idea of the type of annotation work I do. They are always surprised when I tell them] (Personal interview with data annotator for Sapien). This perception of invisibility of their work as annotators mirrors the invisibility of unpaid care work that is automatically assumed by women, especially in a familial context, underscoring Caroline Criado-Pérez's observation that "invisible women are everywhere. They're doing work that isn't counted, that isn't valued, but that is absolutely essential" (Criado Pérez, 2019). Annotation has thus become a hidden extension of global AI infrastructure, just as domestic work sustains the household but goes uncounted. Criado Pérez repeatedly references how "the unpaid care economy" is excluded from GDP and other official measures, erasing the primary work women do to sustain households. We can consider data annotation gigs picked up in support of primary household earnings to parallel this unpaid care economy. The "male-unless-otherwise" design of labor systems normalizes women's overrepresentation in precarious forms of work while rendering their contributions nearly invisible.

Given the lack of knowledge and awareness of data annotation as a legitimate form of work, as well as the growing need for women in Latin America to take on second or third jobs due to inability to buy basic household items like food and medicine because of economic decline, women find themselves pushed toward jobs that embody the flexibility paradox, supposedly allowing them to care for members of their family at the same time as getting paid. We can predict a growth in this area, specifically for workers who need flexibility, like women, migrants without papers, or populations with disabilities, and will be tracking as the need for human data annotators with Spanish and Portuguese skills grow.

**The "Myth of Flexibility" in Platform Labor**

Media has recently taken to describing the work of data annotators as a type of "army of humans" or rank of "foot soldiers" (Lunden, 2025; Dzieza, 2025), echoing themes of Marx's concept of the industrial reserve army, or a pool of surplus labor mobilized during periods of capitalist demand (Marx, 1867). While this type of description parallels the capitalist hunger displayed by frontier model companies as of late, raising multi-billion dollar rounds, it ignores the very human aspect of training artificial intelligence, in which workers are subjected to poor conditions, unpaid work, and treatment akin to invisible domestic or care laborers.

While TechCrunch monikers assume an air of dignity, in reality, annotators are relegated to a secondary, invisible labor market defined by instability, low pay, and high turnover. Rather than the "foot soldiers" that allow AI to function, these workers correspond more to Guy Standing's notion of the "precariat," in which annotators occupy positions without predictability, benefits, or voice, often entirely invisibilized by the end product: artificial intelligence models (Standing, 2011).  In fact, in the face of annotator mobilization for proper working conditions and classification, frontier models lay off droves of their "precariat" to avoid legal battles (Bansal, 2025). For Latin American women, the precariat is not just an abstract category but a daily lived reality. Platforms reproduce segmentation by restricting access to high-value annotation tasks (e.g., 3D mapping, chain-of-thought reasoning) to workers with prior credentials or corporate affiliations, leaving most women in the low-paid, easily replaceable strata.

While the data annotation power players are largely well-known and originate from the Global North – Scale AI, Appen, MTurk (Amazon) – these larger power players often utilize private outsourcing companies around the globe to recruit workers under the names of CloudFactory (Google, Microsoft, etc.), Remotasks (Scale AI), and Outlier (Scale AI).

In areas like Kenya, Nigeria, and India, these data annotation factories are well-known and prop up the majority of the AI industry in the United States and Canada. Even further, by requiring outsourced workers to sign terms and conditions they may not understand, or legal contracts that limit their ability to speak out publicly, platforms construct a purposeful "circle of invisibility" in order to obfuscate the nefarious pathways and dealings of the AI chain (Rest of World, 2025). There is still relatively little known about the exact experience of workers that are shuffled into these "call-center-like" offices in Latin America, Africa, and East Asia.

A recent investigative report from Africa Uncensored described some of the predatory practices of large AI firms like *Mindrift* in South Africa. Freelancers can find hundreds of thousands of job postings available for tutors and trainers, which, despite the name suggesting a higher required skill level, are little more than gigs in digital sweatshops (Raval & Delfanti, 2024). Even further, while these posts advertise "job" availability (see Mercor, as well), in reality, these opportunities are few and far between gigs that result in "one big waiting game… Many workers… say after onboarding they sit for weeks, even months at a time, with empty dashboards and no tasks to complete," given that most annotation firms like *Mindrift* pay out by task rather than by hour, these job postings quickly come to feel like false advertising (Africa Uncensored, 2025). Just as we see in the high-skilled tech sector labor market, where 50% of jobs posted on platforms like LinkedIn are ghost postings (Economy Media, 2025), flooding the market with ghost postings points to an economic slowdown and a permanent change in the future of work.

Paired with ghost job listings and long hours in digital sweatshops, these factory-like headquarters often keep their workers at their computers for incredibly long hours, without access to bathroom breaks or other benefits. This is without mention of the types of content these workers are often exposed to, causing longer-term psychological trauma (Africa Uncensored, 2025). Unsurprisingly, there are fewer women who end up taking these in-person data annotation gigs, namely due to their physical requirements.

The lack of transparency, clarity, and also confusion around subsidiary annotation firms for frontier model companies is by design – the supply chain is deliberately constructed to be hard to map; main power-players like *Scale* use third-party vendors with different names. Inherently, annotating data points reveals much about the systems themselves being developed, and this confused supply chain is what protects this data and the models it trains from being revealed. Data annotation firms often utilize algorithmic distribution of tasks to taggers based on their skill level, demographic, and time on platform. Just as machine learning algorithms can cause widespread bias in LLMs (Blodgett et al., 2024), machine learning algorithms utilized to score certain workers higher in reputation than others have no checks and balances and risk automatically restricting work to a very specific subsector of people.

Several nonprofit watch groups now monitor the precarity of digital gig work. A recent study released by the Fairwork Institute analyzes several different factors for safeguarding basic standards of fair work for data annotators globally: fair pay, fair conditions, fair contracts, fair management, and fair representation. While the myth that gig work provides the flexibility to work from home, bolsters main income with secondary salaries, and allows you to "be your own boss," the experience of data annotators themselves is often far from "flexible" or "liberating." Rather, annotator work arrangements are frequently characterized by:

> uncertain relations, including problems of low and non-payment, fierce competition resulting from an oversupply of labour, long working hours, risks and harm resulting from dangerous tasks (for example, tasks involving exposure to distressing and/or violent content), lack of transparency in management systems (usually operated by automated, algorithmic means), and difficult dispute regulation processes, which often shift the balance of power towards clients (Fairwork Foundation, 2025, p. 6).

One only needs to investigate the treatment of Facebook's content moderators prior to their elimination and replacement with community notes, or Scale AI's tagger exposure to harmful content like bestiality, workers who often refer to their roles as the "front-line soldiers" of the internet. In a case between Scale AI and its annotators:

> Plaintiff Schuster was also a "Reviewer for the Flamingo Safety Project." She was required to view Prompts and images dealing with extremely distressing subject matter like suicide, child abuse, sexual predation, and bestiality. She was tasked with marking these Prompts as either 'Safe' or 'Unsafe' (Schuster et al. v. Scale AI Inc. et al., 2024).

The treatment of these workers has since become so well known that independent filmmakers have started to dramatize their experiences (see *American Sweatshop*, 2024).

**Shifting Demand and the Future of Annotation**

At first glance, many cursory articles published in personal substacks or on LinkedIn may miss what powers our Artificial Intelligence Infrastructure, glossing over that an entire supply chain of workers, human intelligence, data, chips, processors, compute, and electricity is what allows AI to be functional and useable for, let's say, your daily email composition or to-do list. NY Mag emphasizes this idea of the Artificial Intelligence supply chain in a recent article about human intelligence:

> There's an entire supply chain… The general perception in the industry is that this work isn't a critical part of development and isn't going to be needed for long. All the excitement is around building artificial intelligence, and once we build that, it won't be needed anymore, so why think about it? But it's infrastructure for AI. Human intelligence is the basis of artificial intelligence, and we need to be valuing these as real jobs in the AI economy that are going to be here for a while (Dzieza, 2025).

Unveiling the guts that allow models like ChatGPT, Gemini, and Claude to function is essential to understanding the economics of gig work and the direction that model inference is headed.

DeepSeek's public debut in the West in March 2025 caused waves, especially among the data annotation community. Leading annotators, staff, and investors alike to wonder, is this the end for data annotation? For reference, DeepSeek's debut suggested that the model had trained itself upon a small, high-quality dataset paired with reinforcement learning (DeepSeek, 2025), producing a model comparable to OpenAI-o1-1217, which in comparison was trained on many millions of annotated data points. Many assumed data annotation was not long for this world: investors backed out of funding rounds, the market surrounding this supply chain crashed (see Nvidia), and companies like Scale worried about their longevity.

Yet, after the shockwaves of initial reporting faded away, the data annotation community started to see a shift. The conclusion was that data annotation was not going away yet, but there would be a heavier value placed on skills and complex data with higher payouts. This meant tapping into people with multiple language capabilities, doctors, PhDs, and coders alike to provide high-level chain-of-thought reasoning. Companies like Centaur AI, founded by PhDs in Collective Intelligence from universities like MIT, were particularly positioned to succeed given these developments and their recognition in the medical industry as an academically trusted annotation source.

What we can take away from this is not only that we will see a transformation in the type of work annotators are doing—from quantity toward quality—in turn, shaping the workforce and placing more value on highly skilled annotators. This means that many of the ground-level annotators conducting basic human reasoning like "Is this image of a cat?" or "Is the bounding box around the proper object?" that got artificial models off the ground in the first place, will be displaced from their current work. We can already see this with a decrease in access to general tasks at Scale AI, which was recently partially acquired by Meta and went through a layoff of 400 employees. Taskers on Outlier, Scale's subsidiary, Spanish-speaking community board repeatedly ask when new tasks will become available to no avail: "Alguna novedad? Alguien sabe si hay proyectos activos? Veo todo detenido." [Anything new? Does anyone know if there are active projects? I see everything paused.] (Outlier website, 2025).

There is a flipside to this movement toward skilled work, however, even if it is solely in ideology. As discussed, data annotation has consistently either not been seen or been viewed as factory line-level work that anyone with basic reasoning skills could conduct. Much of this framework explains why the data annotation sector has bifurcated across gender and geography. This has led companies and society alike to act not only as if the annotators that built the models used every day don't exist but also to pay them wages so low that these posts can only be considered indentured servitude.

We can ride the wave of reframing annotation as "skilled labor" in the wake of DeepSeek, promoting annotation as a high-value job and one of the only types of work that supports an economy that is projected to hit $4.8 trillion by 2033 (UNCTAD, 2025). If reframed, even in ideology, as skilled labor, annotators can argue for further protections than what they are currently promised and feel a higher level of dignity in their work. Not only this, but annotation can come to be seen in a similar light to Uber and Lyft, which are now heavily regulated by state, local, federal, and international governments alike (U.S. Department of Labor, 2025).

**Regulatory Interventions and Platform Accountability Mechanisms**

The distinction between freelancers, contractors, and gig workers must be situated within the decades-long fissuring of employment, in which platforms have made it cheaper to externalize volatility and "govern" work remotely from the Global North. The governance of annotation reflects what Crane et al. (2017) describe as a "governance gap" in global value chains. Platforms headquartered in the United States and Europe contract labor from Latin America but remain largely outside local regulatory reach.

The global nature of data annotation work raises urgent questions of jurisdiction and accountability. The Global Value Chain theoretical framework, starting with Hopkins and Wallerstein's coining of the global commodity chain concept, or the "network of labor and production processes whose end result is a finished commodity," predicted changes to international divisions of labor with the rise of globalization (Hopkins & Wallerstein, 1986). With each passing year and technological advancement, these commodity chains have shifted toward more sophisticated types of labor-value commodity chains, in which the extraction of surplus value and labor exploitation from the Global South remains central. Gary Gereffi recognized as early as 2005 a shift of production and "a very large and growing proportion of the workforce… located in developing economies" (Gereffi, 2005).

In practice, this refers to the fact that most annotators in Latin America perform tasks for firms headquartered in the United States, Canada, and England, yet regulatory protections remain rooted in national frameworks that rarely extend across borders. This asymmetry has created what labor scholars describe as a "governance gap" in the platform economy (Crane et al., 2017). This governance gap can be linked to the continuance of forced labor in domestic and international supply chains.

International labor institutions seek to address this gap. At the international level, the ILO has called for extending protections to all workers, and until recently, this compliance was only voluntary (Berg, 2019). The Fairwork Foundation benchmarks platforms on fair pay, conditions, contracts, management, and representation, though these remain primarily advocacy tools. The

EU Platform Work Directive (2024) has gone furthest by creating a presumption of employment and requiring algorithmic transparency. Latin America lacks such binding frameworks, as regional bodies like Mercosur and the Andean Community have not established standards for digital labor.

As recently as June 5, 2025, the ILO gathered to establish binding labor standards for the platform economy, with 187 member-states in attendance at the 113th conference in Geneva. This ILO agreement arose as the reports of poor treatment of platform workers and tax and social security evasion strategies of platform companies grew in size. While draft versions of this agreement have circulated, representatives have yet to finalize aspects regarding misclassification of workers, use of data, and what kind of recommendation and/or convention could be held at the 2026 convention, providing more substantive protections for workers at the bottom of the AI supply chain. Because the ILO is an intergovernmental body, calling a Convention would signify agreement on minimum labor standards for workers, establishing a legally binding and morally significant set of recommendations across a multitude of nations.

Diving further into the ILO's investigation and reporting on platform work, Janine Berg's research on crowdworking and on-demand economies highlights how firms use digital tools to shift risk onto workers through outsourcing and platform intermediation, pushing work and its risks onto the "Majority World." Technology, rather than replacing monopsony power, scales and obscures it through algorithmic management, fissuring, and misclassification. Berg shows that deteriorating job quality follows from firms using digital tools to shift risk onto workers through outsourcing and platform intermediation. Technology does not replace classic sources of monopsony power but rather scales and obscures them through algorithmic management, fissuring, and status misclassification.

She advocates for three pillars of policy change: extending protections and collective rights to all workers; reducing working time to recognize unpaid care; and ensuring participatory, transparent, and privacy-preserving technology design (Berg, 2019). These frameworks help the international community better understand hierarchies of labor. Berg maps labor starting from "core standard jobs" (9-5 in-person workplaces), to non-standard jobs (freelance, gig work, domestic labor), to an outer ring of unpaid "heteromated" labor in which users perform valuable but uncompensated tasks. Unlike service sectors, which tend to be highly regulated at a national level, even providing pay for training hours, Berg calls our attention to the unpaid digital labor that annotators undertake in the form of onboarding, screening, reputation building, and "leveling up."

Watchdog organizations like the Fairwork Foundation and the European Union's I-CLAIM have developed a global benchmark for "fair platform labor" based on five pillars: fair pay, fair conditions, fair contracts, fair management, and fair representation. More recently, the foundation launched a "Fairwork AI Supply Chain Certification" for companies that work hand in hand with the organization to provide transparency in their supply chain. While influential in debate, these remain advocacy instruments rather than enforceable regulation. At international labor convenings, the very leaders of Fairwork lamented that these certifications unfortunately do not sway larger companies like OpenAI or Meta to change vendors, given that their ultimate focus is on the bottom line.

Nationally, U.S. regulations on misclassification (U.S. Department of Labor, 2025) and state-level rulings, such as Minnesota's minimum pay for app-based workers, set precedents but do not extend transnationally. In Latin America, fragmented labor codes often exclude freelancers, creating a regulatory vacuum that enables platforms to exploit global labor arbitrage. Recent legislation in the United States has tightened the prerequisites for classification as an employee or independent contractor (U.S. Department of Labor, 2025). The Department of Labor notes that "misclassification of employees as independent contractors may deny workers minimum wage, overtime pay, and other protections." Although this does not immediately redefine standards for gig workers in the U.S., it increases the burden on companies that classify workers as independent contractors without extending benefits.

At the state and local level in the United States, Minnesota's 2025 ruling set minimum pay for app-based workers and increased scrutiny of misclassification across industries. Nebraska, by contrast, voted 33–15 to lock gig drivers in as contractors (Latimer, 2025).

At the regional level, the European Union has advanced the furthest with its 2024 Platform Work Directive, which creates a presumption of employment for many platform workers and requires algorithmic transparency in task allocation. No comparable framework exists in Latin America, where fragmented labor codes exclude many digital freelancers. The Andean Community and Mercosur, South American trade blocks with influence on the economic market in the region, have yet to establish common standards on platform labor, leaving gig workers vulnerable to global labor arbitrage. As recently as September 5, 2025, trade unions began to raise concerns about lack of protection for workers in the European Union-Mercosur trade agreement currently in draft form. As we have seen in the case of Toloka, a transatlantic agreement without explicit protections only threatens to deepen exploitation of workers living in more precarious economic situations (Latin America-Spain relations, for example). Given that the EU-Mercosur's own joint impact assessment on employment sustainability cites that taskers performing platform work are subject to "slave-like working conditions," cross-border agreements that fail to provide any significant real wage improvement or benefits seem insufficient for approaching treatment of taggers centralized in precarity in Venezuela, Colombia, and Peru.

**Conclusion**

For decades, fissuring reporting on employment created a false positive narrative around job growth and the potential for alternative labor markets like capital and labor platforms to release low-wage workers from their shackles to larger companies. Platform work demonstrates yet another successful attempt by the likes of Meta, OpenAI, Google, Microsoft, and other conglomerates to lower their costs by externalizing volatility and risk and monitoring data work remotely from lofty headquarters in San Francisco.

This risk externalization reflects what Crane et al. (2017) describes as a "governance gap" in global value chains. Platforms headquartered in the United States and Europe contract labor from Latin America but remain largely outside local regulatory reach, avoiding both local labor laws as well as global regulation.

For annotators, this regulatory vacuum means global AI firms can exploit the lack of enforceable cross-border protections while drawing surplus labor from the Majority World. Women, already burdened with caregiving responsibilities and concentrated in lower-skilled tasks, face the most severe consequences of this governance gap.

The stakes of this moment extend far beyond individual workers or even national economies. As artificial intelligence systems increasingly shape global decision-making, from healthcare diagnostics to criminal justice algorithms, the labor that trains these systems becomes a matter of democratic governance. When annotation work remains invisible, precarious, and concentrated among the most vulnerable populations, we embed inequality into the very foundation of our technological infrastructure. The biases, shortcuts, and survival strategies that emerge from exploitative working conditions do not disappear when the AI model is deployed: they scale globally, affecting millions who will never know that their loan rejection, job application screening, or medical recommendation was shaped by workers earning less than minimum wage in conditions that deny them basic dignity.

Without intervention, annotation will continue to operate as an unregulated labor reservoir, perpetuating colonial patterns of resource extraction into the digital age. With intervention, it could be reframed as skilled, essential labor deserving protections and dignity. The choice we make today about how we value this work will determine not only the livelihoods of millions of annotators but the democratic potential of the technologies that increasingly govern our lives.